  \documentclass[multi]{cambridge6Atight}
  \usepackage{natbib}

  \usepackage{rotating}
  \usepackage{floatpag}
  \rotfloatpagestyle{empty}

 \usepackage{amsmath}
  \usepackage{amsthm}
  \usepackage{graphicx}
  \usepackage{mathptmx}

  \def\teff{\ensuremath{T_\text{eff }}}
  \def\msol{\ensuremath{M_\odot}}
  
  \def\omcrit{\ensuremath{\Omega_\text{crit}}}
  \def\geff{\ensuremath{\vec{g}_\text{eff}}}

  \usepackage{makeidx}
  \makeindex

 \copyrightline{Reprinted from \textit{The Impact of Binaries on Stellar Evolution}, Beccari G. \& Boffin H.M.J. (Eds.), \copyright\ 2018 Cambridge
    University Press.}

\begin{document}


  \alphafootnotes
   \author[C. Georgy and S. Ekstr\"om]
    {Cyril Georgy\
    and Sylvia Ekstr\"om\footnotemark}
  \chapter{Massive Star Evolution: Binaries as Two Single Stars}

  \footnotetext[1]{The authors acknowledge support from the Swiss National Science Foundation (project number 200020-172505). CG acknowledge PRACE for awarding access to resource MareNostrum 4  based in Spain at Barcelona Supercomputing Center. The support of David Vicente and Janko Strassburg from Barcelona Supercomputing Center, Spain, to the technical work is gratefully acknowledged.}
  \arabicfootnotes

  \contributor{Cyril Georgy
    \affiliation{Department of astronomy, Geneva University, Maillettes 51 - Sauverny, 1290 Versoix, Switzerland}}

  \contributor{Sylvia Ekstr\"om
    \affiliation{Department of astronomy, Geneva University, Maillettes 51 - Sauverny, 1290 Versoix, Switzerland}}

 \begin{abstract}
Binary stars are of course more than two stars, but they are also at least two stars. In this chapter we will review some aspects of the physics governing the evolution of single massive stars. We will also review the uncertainties of key physical ingredients: mass loss, rotation and convection.
 \end{abstract}

\section{The basics}
\label{Georgy_basics}
Massive stars\index{massive star} are stars more massive than about $8-10\,M_\odot$, that are bound to die in the explosion of a supernova or collapsing directly into a black hole \citep{Heger2003b,Jones2013a}. Their life is marked by the succession of all burning stages up to silicon burning.

The modelling\index{modelling} of massive stars is done by the mean of one-dimension (1d) stellar evolution codes (see below), which solve the four equations of stellar structure \citep[see][for a classical textbook]{Kippenhahn1990a}, coupled with a network of nuclear reactions providing the energy generation rate inside the stellar model, and the time evolution of the chemical content of the star. On top of that, various effects can be accounted for, such as rotation\index{rotation} or mass loss\index{mass loss}, particularly important in the context of the evolution of massive stars.

An important aspect of massive star evolution that became prominent during the past decade is that a significant fraction of massive stars are found in multiple systems. Large surveys in different environments suggest that $50\%$ to $70\%$ of all massive stars \citep{Sana2012a,Sana2013a} are in binary (or multiple) systems where components are sufficiently close to interact at least once during their lifetime. Evolution of massive stars in multiple systems can be very different from single star evolution \citep[][among others]{vandenHeuvel1975a,Vanbeveren1991a,Podsiadlowski1992a,Vanbeveren1998a,Eldridge2008a,deMink2013a}. Taking multiplicity into account is thus extremely important when dealing with massive star populations.

It is however important to keep in mind that it is utopian to model correctly multiple stellar systems without being able to accurately describe the evolution of a single star. In this chapter, we discuss uncertainties of key physical processes taking place in single star models and their consequences. These shortcomings should be kept in mind by anybody using results from stellar evolution computations or from population synthesis.

\section{Mass loss\index{mass loss}}
\label{Georgy_massloss}

Mass loss\index{mass loss} is a very important ingredient to be included in stellar evolution code when dealing with massive stars. At solar metallicity, a $15\,M_\odot$ star loses $2$ to $3\,M_\odot$ during its entire life (but only a few tenths of a solar mass during its main sequence). At the other extremity of the mass spectrum, a $120\,M_\odot$ star loses up to $100\,M_\odot$ during its lifetime, including several tens during the main sequence \citep[see e.g.][]{Ekstrom2012a}. The situation is even more extreme in the case of very massive stars, which lose most of their mass in the course of their life \citep{Yusof2013a}.

\begin{figure}
\includegraphics[width=.6\columnwidth]{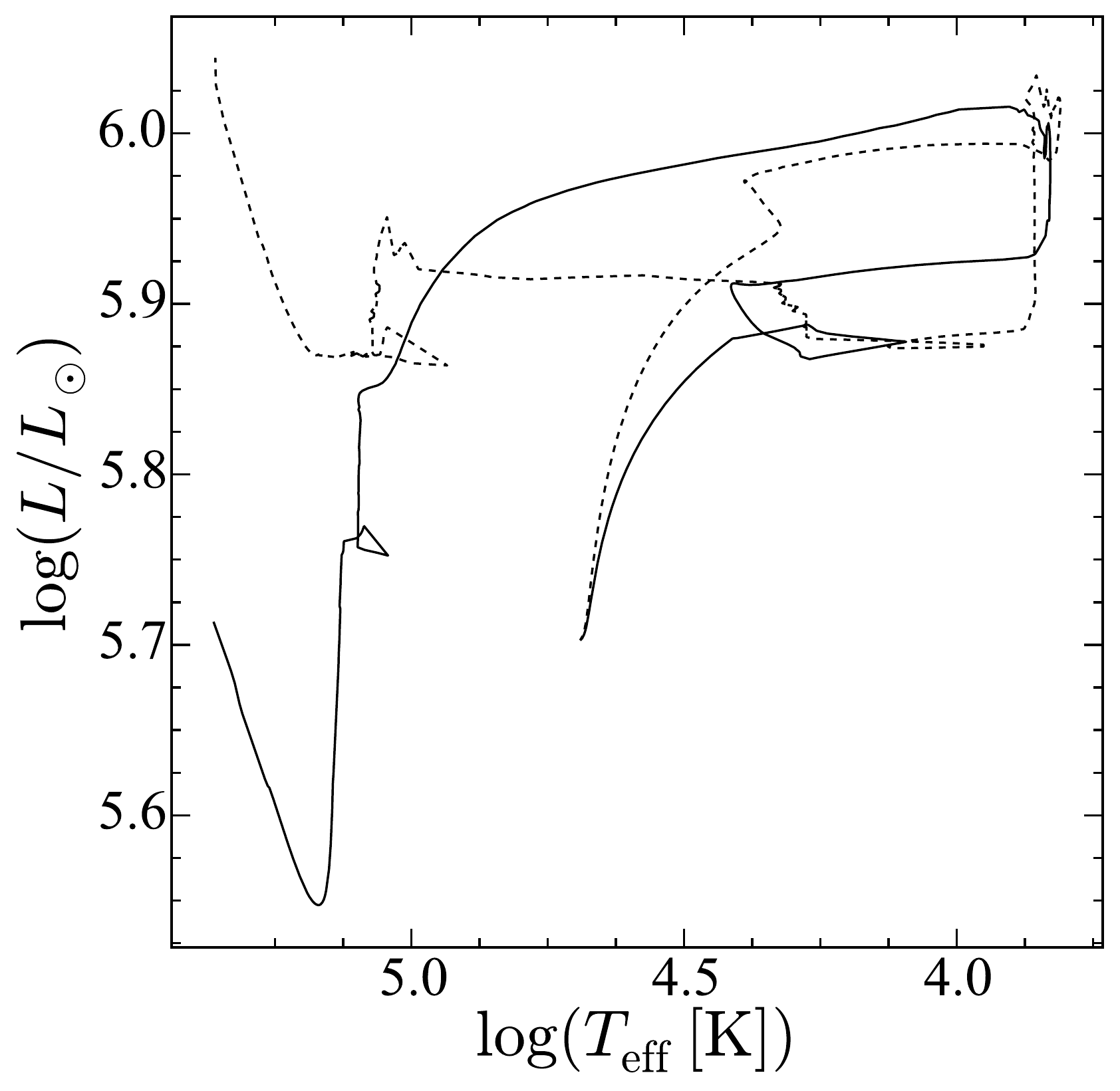}
\caption{Tracks in the Hertzsprung-Russell diagram of a $60\,M_\odot$ model computed with two different prescriptions for the mass-loss rates during the main-sequence: \citet[solid line]{Vink2000a} and \citet[dotted line]{Kudritzki2000a}. The impact on the tracks is clearly visible, due to the much higher rates with the \citet{Vink2000a} prescription.}
\label{GeorgyFig:MassLossCompa}
\end{figure}

During the main sequence phase, winds from massive stars are in general described in the framework of the radiation-driven wind theory \citep[``CAK'' theory, see][]{Lucy1970a,Castor1975a}. It has been improved over the years to accommodate for supplementary processes such as multiple scattering, wind clumping, etc. \citep[see the review by][]{Puls2008a}. It has lead to the computation of theoretical mass-loss rates that are extensively used in stellar evolution calculations \citep{Vink2000a,Vink2001a}. However, the face-values of the obtained mass-loss rates computed with this recipe seems to disagree by a factor 2-3 with mass-loss rates obtained observationally at different wavelengths \citep[e.g.][]{Najarro2011a,Surlan2013a,Rauw2015a}. This uncertainty on the mass-loss rates that should be adopted in stellar evolution models has a major impact on the results of such simulations. As an illustration, Fig.~\ref{GeorgyFig:MassLossCompa} shows tracks in the Hertzsprung-Russell diagram for two $60\,M_\odot$ models with different prescriptions for the computation of the mass-loss rates on the main-sequence: a model using \citet[solid line]{Vink2000a} and one using \citet[dotted line]{Kudritzki2000a}. Mass-loss rates during the advanced stages are kept the same \citep[see also][]{Keszthelyi2017b}. It shows that a difference of a factor of two in the mass-loss rates during the main sequence modifies completely the late evolution and the end-point just before the supernova.

For the most massive O\index{O star} and Wolf-Rayet (WR)\index{WR star} stars, a dependence of the mass loss rates on the Eddington factor $\Gamma_\text{Edd}=\frac{\kappa L}{4\pi cGM}$ seems to appear when they reach large $\Gamma_\text{Edd}$ \citep{Vink2006a,Vink2011c,Grafener2011a,Bestenlehner2014a}. This dependence is around ${\Gamma_\text{Edd}}^2$ for O-stars with $\Gamma_\text{Edd}<0.70$, and around ${\Gamma_\text{Edd}}^4$ for WR stars with $\Gamma_\text{Edd}>0.70$.

Mass loss rates usually scale with the metallicity\index{metallicity}, but the scaling to adopt is not yet completely clear. The $Z$ dependence has been estimated initially to be 0.4 by \citet{Abbott1982a}, and then between 0.5 \citep{Kudritzki2002a} and 0.85 \citep{Vink2001a}. \citet{Mokiem2007b} find the empirical scaling deduced from observations of early B-type and O-type stars to be 0.83 for a smooth wind solution, or 0.72 when a correction for clumping is applied.

During the red supergiant\index{red supergiant} phase, the mass-loss rates are even less well known. Observational determinations exhibit a very strong scatter, with variations over more than 2 orders of magnitude at a given luminosity \citep{Mauron2011a,vanLoon2005a,Beasor2016a,Beasor2017a,Georgy2017d}. As the winds from red supergiants are still not well understood from a theoretical point of view, the current prescriptions used in stellar evolution code are fits to these observational data. Questions such as: ``in what kind of star, when and how long do the heavy mass-loss episodes occur? What is the average mass-loss rate at a given evolutionary stage?'' need to be answered in order to correctly model the late stages of stellar evolution.

In case the mass-loss rates are best described by the prescriptions leading to rates in the low range of the observations, stars up to about $30\,M_\odot$ end their life as a red supergiant. In such case, explanations to the so-called ``red supergiant problem'' \citep{Smartt2009a} have to be found. Strong extinction of the supernova event due to the dust in the strong winds of the more massive red supergiants \citep{Walmswell2012a} or direct collapse into a black hole \citep{Adams2017a} have been suggested. In case average mass-loss rates during the red supergiant phases are underestimated in usual prescriptions, it allows the star to evolve back toward the hot side of the Hertzsprung-Russell diagram \citep{Vanbeveren1998a,Georgy2012a,Meynet2015a}. In this context, it is also interesting to note that a theoretical explanation of the pulsating properties of a class of variable blue supergiants, the ``$\alpha$-Cygni'' variables, seems to require that these stars have lost their hydrogen-rich envelope through a strong mass-loss episode, possibly during a previous red supergiant stage \citep{Saio2013a}.

\section{Internal mixing processes\index{mixing}}
\label{Georgy_mixing}

Mixing processes\index{mixing} are any physical process able to transport chemical elements or angular momentum from one place to another inside a star. They are important as they can considerably affect the way a star evolves, as well as its surface (and observable) properties. Historically, the first of such process included in the modelling of stars was convection \citep[even before people started considering heat transfer by radiation, see][]{Eddington1926a}. Rotation is another such process and has been studied very early during the 20th century \citep{vonZeipel1924a,Eddington1925a,Endal1976a}. In this chapter, we focus only on these two processes \citep[each of them would still require a whole book for an in depth description, like][for rotation]{Maeder2009a}, and we will not consider other mixing processes such as radiative diffusion, gravitational settling, internal waves, magnetic field, or tides, among others.

\subsection{Convection\index{convection}}
\label{Georgy_convection}
The treatment of convection in classical 1d stellar evolution codes is a long standing problem of stellar evolution \citep[e.g.][]{Arnett2015a}. In most of stellar evolution codes (e.g. the Geneva Stellar Evolution code, \citealt{Eggenberger2008a}; MESA, \citealt{Paxton2011a,Paxton2013a,Paxton2015a,Paxton2017a}; STERN, \citealt{Petrovic2005a}; KEPLER, \citealt{Heger2000a}; FRANEC, \citealt{Chieffi2013a} for the most widely used code in the massive star community), it is a two steps process: 1) determine the boundaries of the convective zones, often using the ``Schwarzschild'' or ``Ledoux'' criterion \citep[e.g.][]{Kippenhahn1990a}, and possibly extend them empirically to account for the ``overshoot'' \citep{Maeder1975a} which cannot be predicted by the linear perturbation theory used to elaborate the stability criteria; 2) compute the thermal gradient inside the convective zone, often using the ``mixing-length theory'' (MLT) or one of its derivative \citep{Boehm-Vitense1958a}. This was devised before the publication in the west of Kolmogorov's work on turbulent cascade \citep{Kolmogorov1941a,Kolmogorov1962a}, and may be inconsistent with that now well-established theory \citep[see also][]{Arnett2009a}.

There are a variety of ways of implementing convection in 1d stellar evolution codes, following theoretical works or guided by hydrodynamics simulations \citep{Zahn1991a,Freytag1996a,Meakin2007a}. Recent works have shown how sensitive the outputs of 1d stellar modelling are to the detailed implementation of convection \citep{Martins2013b,Georgy2014a}.

\begin{figure}
\includegraphics[width=.7\columnwidth]{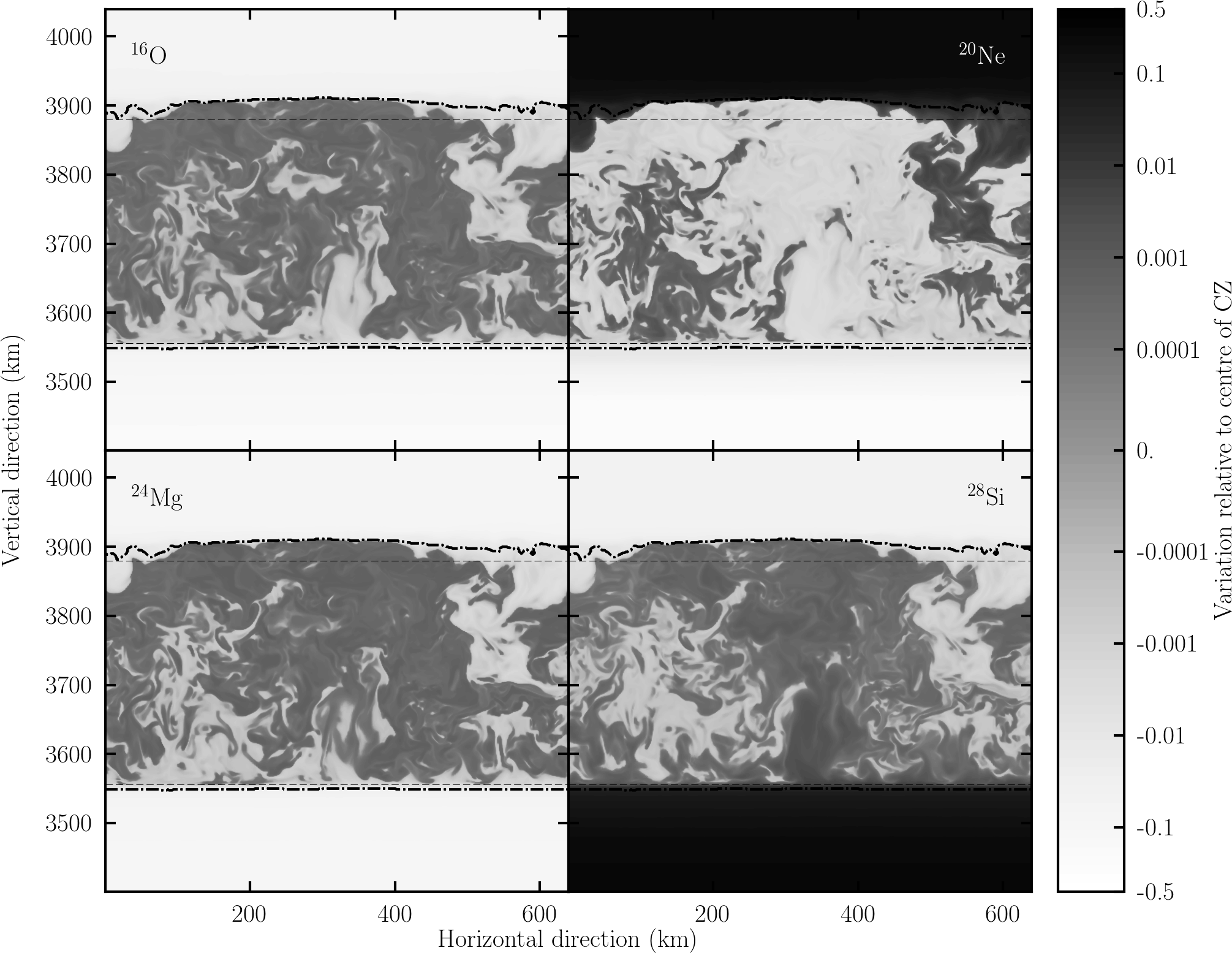}
\caption{Abundance of $^{16}\mathrm{O}$ (top-left panel), $^{20}\mathrm{Ne}$ (top-right panel), $^{24}\mathrm{Mg}$ (bottom-left panel), and $^{28}\mathrm{Si}$ (bottom-right panel) relative to the average value in the middle of the convective zone. Black and dark greys indicate overabundances while white and light greys indicate underabundances. The figure shows a 2D vertical slice through the 3D computation domain. The thin dashed line indicates the position of the convective zone in the corresponding 1d model, and the thick dot-dashed line shows the convective boundary in the 3d simulation.}
\label{GeorgyFig:NeBurning}
\end{figure}

In this context, hydrodynamics simulations\index{hydrodynamics} of convection in stellar interiors are very important, as they allow one to study the behaviour of the flow inside the convective regions from first principles. With the rapid increase of computational resources during the past years and the development of efficient parallelised hydrodynamics code, it has become possible to perform such simulations in different regimes: convective envelopes of cool stars \citep[e.g.][]{Freytag2008a,Chiavassa2009a,Viallet2013a,Magic2013a}, intermediate helium-burning shell in intermediate mass stars \citep{Herwig2006a,Woodward2015a}, or late burning stages of massive star life, such as carbon-burning shell \citep{Cristini2017a}, oxygen-burning shell \citep{Meakin2007a,Muller2016a,Jones2017a}, or silicon-burning shell \citep[e.g.][]{Couch2015a}. These simulations provide very important informations about the physics of the convective boundary (shape, location, motion, boundary mixing) and the interaction between the nuclear burning and the convection (see Fig.~\ref{GeorgyFig:NeBurning}, where the chemical abundances of $^{16}\mathrm{O}$, $^{20}\mathrm{Ne}$, $^{24}\mathrm{Mg}$, and $^{28}\mathrm{Si}$ with respect to the centre of the convective zone are shown in the case of the simulation of a neon-burning shell). The results of such simulations can be used to deduce prescriptions for the convective boundary mixing \citep{Freytag1996a,Herwig2000a,Jones2017a}. Using specific averaging techniques \citep[``RANS'' approach, see e.g.][]{Viallet2013a,Mocak2014a}, they are also useful in providing a new theoretical framework and new algorithms for the treatment of convection in 1d stellar evolution code \citep{Arnett2015a}.

The need for a new treatment of convection and convective boundary mixing is observationally motivated by the asteroseismic data available for main-sequence B-type stars \citep{Moravveji2015a}, or from nucleosynthesis considerations at different production sites \citep{Battino2016a,Choplin2017a}. The modelling of convection has also a deep impact on the final structure of the star, and can considerably affect its final fate by changing its compactness \citep[supernova explosion or not, and the type of remnant left behind, see][]{OConnor2011a,Sukhbold2014a,Davis2017a}.

\subsection{Rotation\index{rotation}}
\label{Georgy_rotation}
Rotation has two different ways of influencing stellar evolution. 1) It creates a deformation of the star's shape such that the stellar characteristics become dependent on the colatitude considered, affecting the stellar parameters deduced from observations. It also induces an anisotropy in the mass lost by the star. 2) It triggers an internal mixing that transports chemicals and angular momentum, and induces many deviations from the standard non-rotating evolution. We detail both effects below.

\subsubsection{Surface deformation}
The deviation from spherical symmetry is proportional to the rotation ratio $\omega=\Omega_\text{surf}/\omcrit$\footnote{We use here $\omcrit=\sqrt{\frac{8}{27}\frac{G\,M}{R_\text{pol,crit}^3}}$, where $R_\text{pol,crit}$ is the polar radius at the critical velocity.}. In the frame of the Roche model, the equatorial radius is at most $R_\text{eq,crit} = \frac{3}{2}\,R_\text{pol,crit}$, with the reasonable assumption that $R_\text{pol,crit}=R_{\text{pol,}\omega=0}$. An oblateness has been observed by interferometry in the case of the rapidly rotating star Achernar ($\alpha$ Eridani) for example, in agreement with the value of 1.5 \citep{Domiciano2003a,Vinicius2006a,Carciofi2008a}.

Because of the deformation, the effective gravity in the Roche approximation becomes dependent on the rotation rate and the colatitude: $$\geff=\geff(\Omega,\theta)= \left( -\frac{G\,M}{r^2} + \Omega^2r\ \sin^2\theta \right)\ \vec{e}_r + \Omega^2r\ \sin\theta\ \cos\theta\ \vec{e}_\theta.$$
Subsequently, the flux inherits from this $\Omega-\theta$ dependence \citep[e.g.][]{Maeder1999a}: $$\overrightarrow{F}=\overrightarrow{F}(\Omega,\theta)= -\frac{L}{4\pi\,G\,M^\star}\ \geff(\Omega,\theta),$$
where $M^\star = M\ \left( 1- \frac{\Omega^2}{2\pi\,G\,\rho_M} \right)$. According to the Stefan-Boltzman law, we have $F=\sigma T^4$, so the effective temperature also becomes dependent on $\Omega$ and $\theta$ \citep{vonZeipel1924a}:
$$\teff = \left[\frac{L}{4\pi\sigma GM^\star}g_\text{eff}(\Omega,\theta)\right]^{1/4}.$$  \citet{EspinosaLara2011a} have proposed a more general expression, valid also in the case of rapid rotation:
$$\teff = \left(\frac{L}{4\pi\sigma GM}\right)^{1/4}\sqrt{\frac{\tan\vartheta}{\tan\theta}}\ g_\text{eff}^{1/4}$$
where $\vartheta$ is the solution of the equation:
$$\cos\vartheta + \ln\tan\frac{\vartheta}{2} = \frac{1}{3}\omega^2\tilde{r}^3\cos^3\theta + \cos\theta + \ln\tan\frac{\theta}{2}.$$
The latitude dependence of the temperature has been observed in interferometry for the star Altair \citep[$\alpha$ Aquilae, see][]{Monnier2007a} and the stars Alderamin and Rasalhague \citep[$\alpha$ Ceph and $\alpha$ Oph, see][]{Zhao2009a}, with values compatible with the models.
\begin{figure}
\includegraphics[width=.7\columnwidth]{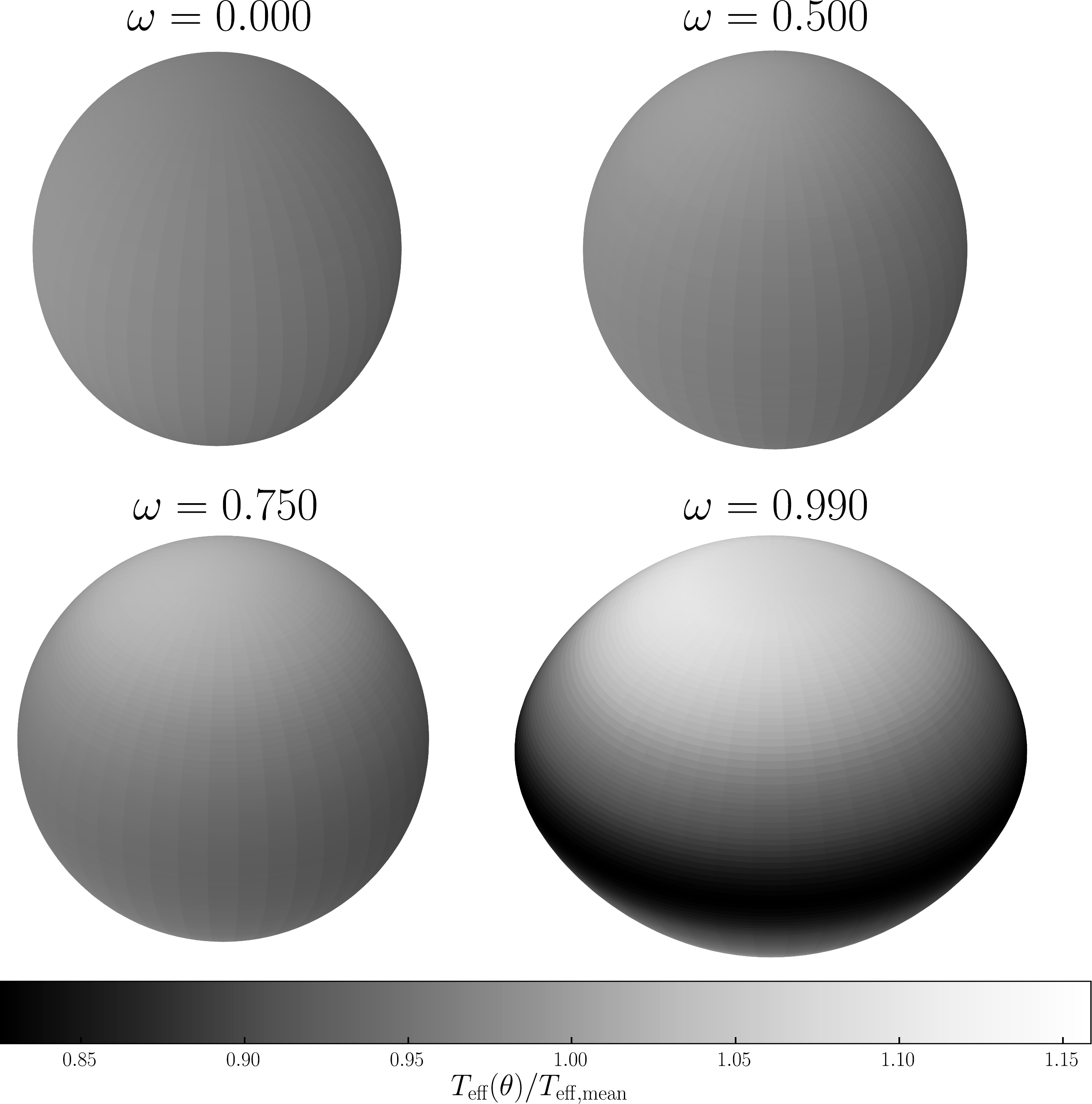}
\caption{Effective temperature of rotating stars relative to the mean \teff for four different rotation rates $\omega=\Omega_\text{surf}/\omcrit$. The $\teff(\theta)$ law is from \citet{EspinosaLara2011a}.}
\label{GeorgyFig:RotSurf}
\end{figure}

Below $\omega=0.7$, the deformation remains insignificant (see Fig.~\ref{GeorgyFig:RotSurf}), but above this value, the angle with which the stars are observed becomes an important parameter. A star observed pole-on will present a higher $L$ and \teff\ than the same star seen with an average inclination. In contrast, a star seen equator-on will appear dimmer and cooler \citep[e.g.][]{Georgy2014b}. The stellar parameters deduced for such stars will be affected.

Hotter poles and cooler equators modify the mass flux ejected by the star, that gets a latitudinal dependence \citep{Maeder2000b}:
$$\frac{\text{d}\dot{M}(\theta)}{\text{d}\sigma} \sim A(\alpha,k)\ \left(\frac{L}{4\pi\ G\ M^\star}\right)^{\frac{1}{\alpha}-\frac{1}{8}} \frac{g_\text{eff}^{1-\frac{1}{8}}}{\left( 1-\Gamma_\Omega(\theta) \right)^{\frac{1}{\alpha}-1}}$$
where $\text{d}\sigma$ is the surface unit, $A(\alpha,k)$ is an empirically determined function of the force multiplier parameters $\alpha$ and $k$ \citep{Lamers1995b}, and $\Gamma_\Omega(\theta)$ is the local Eddington factor taking into account the effects of rotation:
$$\Gamma_\Omega(\theta)=\frac{\kappa_\text{es}L}{4\pi cGM\left(1-\frac{\Omega^2}{2\pi G\rho_\text{m}}\right)}$$
with $\kappa_\text{es}$ the electron-scattering opacity, and $\rho_\text{m}$ the internal average density. According to \citet{Georgy2011a}, for a nearly critically-rotating star, we have $\dot{M}(\text{pol})=3.25\,\dot{M}(\text{eq})$. This anisotropy modifies the angular momentum budget, since the mass loss occurs preferentially in the region of the poles, which removes less angular momentum than what the equatorial regions would do \citep[see also][]{Muller2014a}.

\subsubsection{Transport processes}

In 1D, the transport of angular momentum\index{angular momentum} is, according to \citet{Zahn1992a,Maeder1998a}:
$$\rho \frac{\partial}{\partial t} \left( r^2 \bar{\Omega} \right)_{M_r} =\underbrace{\frac{1}{5r^2} \frac{\partial}{\partial r} \left( \rho r^4 \bar{\Omega} U(r) \right)}_\text{advective term} + \underbrace{\frac{1}{r^2} \frac{\partial}{\partial r} \left( \rho D_\mathrm{v} r^4 \frac{\partial \bar{\Omega}}{\partial r} \right)}_\text{diffusive term}$$
where $U(r)$ is the function such that
the radial component of the meridional circulation is $U(r)\ P_2\cos(\theta)$, and $D_\text{v}$ the diffusion coefficient due to various mechanisms like convection or shear. The transport of chemical species has been shown to be satisfactorily approximated by a diffusion-only process \citep{Chaboyer1992a} and can be expressed as:
$$\rho\frac{\partial X_i}{\partial t} = \frac{1}{r^2}\frac{\partial}{\partial r}\left(\rho r^2\left(D_\text{v} + D_\text{eff}\right)\frac{\partial X_i}{\partial r}\right)$$
with $D_\text{eff}$ the diffusion coefficient due to meridional circulation and horizontal turbulence.

The implementation of rotation in the various codes existing on the market varies a lot from one to another. Some of them use a diffusion-only approximation to express the transport of angular momentum, as for the species. When the advective term is taken into account, three different  expressions for the horizontal turbulence can be used \citep{Zahn1992a,Maeder2003a,Mathis2004a}, and two for the shear turbulence \citep{Maeder1997a,Talon1997a}. \citet{Maeder2013a} proposed also a global diffusion coefficient covering all the various instabilities triggered by rotation (GSF and Solberg-H\o iland instabilities, thermohaline mixing, ...) and the interplay they have one with each others. The various ways of expressing the mixing process induced by rotation leads to large differences between the outputs of different codes, as highlighted by \citet{Chieffi2013a}. A study of the impact the different implementations have on the stellar evolution is presented in \citet{Meynet2013a}.

\subsubsection{Effects on the evolution}
\begin{figure}
\includegraphics[width=.6\columnwidth]{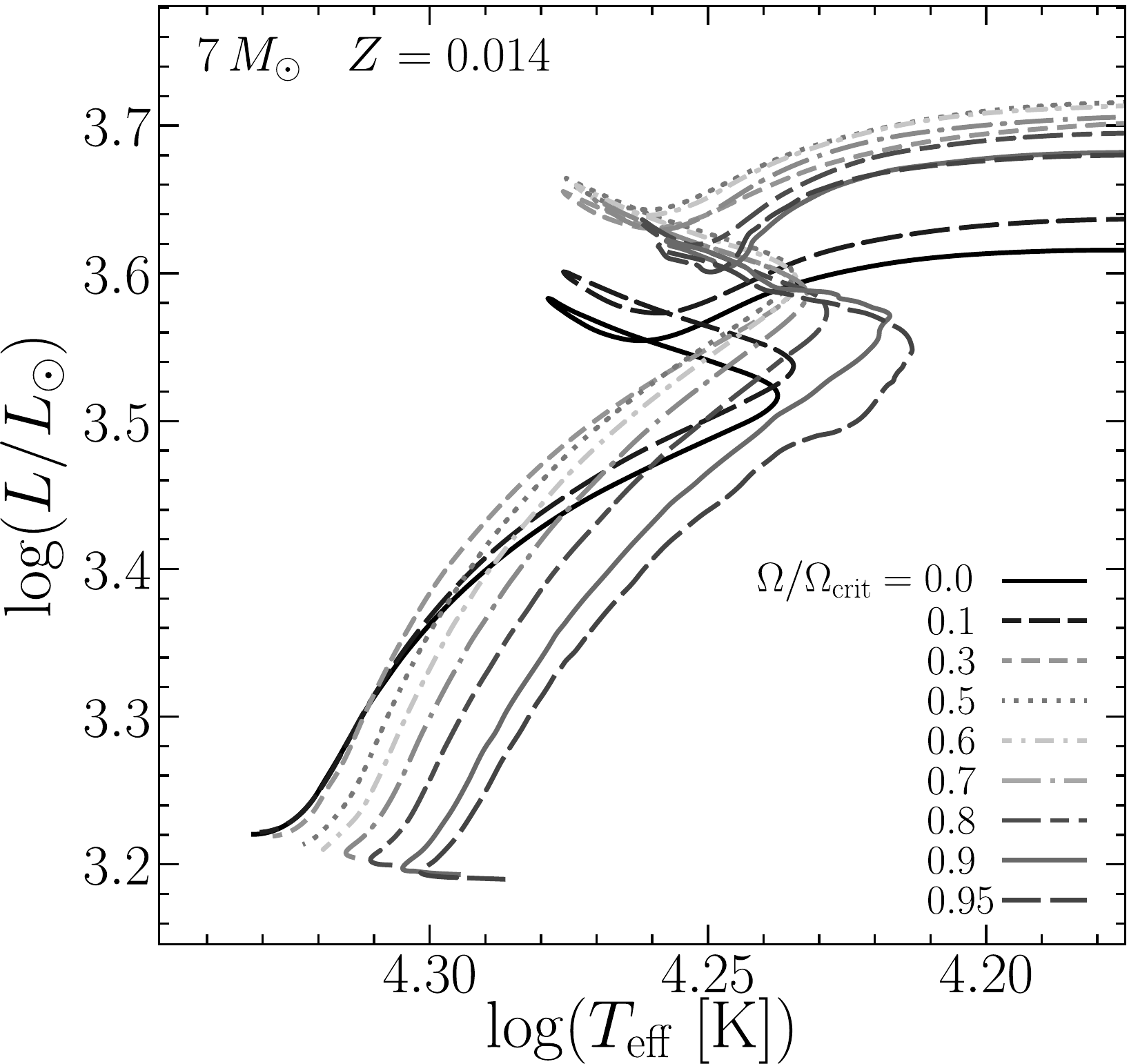}
\caption{Main sequence tracks in the Hertzsprung-Russell diagram for models of $7\,\msol$ with various initial rotation rates. \citep[Models from][]{Georgy2013a}.}
\label{GeorgyFig:MS7M}
\end{figure}

Rotational mixing brings fresh fuel into the core, allowing it to grow further, hence resulting in longer lifetimes and a higher luminosity. However, the behaviour of a star is not monotonic with increasing rotation. At the beginning of the main sequence, the hydrostatic effects dominate and the star presents a slightly lower $L$ and \teff. As the evolution proceeds, the mixing induces a slower shrinking of the core, increasing the luminosity, while the \teff\ decreases because of the modified chemical structure. Note that the effects of mixing are dominant for low and average rotation, while the rapid rotators show a composite behaviour between strong hydrostatic effects and mixing effects. The brightest star are therefore not the fastest ones, but the slightly-above-average rotators (see Fig.~\ref{GeorgyFig:MS7M}).
In some extreme cases, the mixing during the main sequence is so strong that the star evolves quasi chemically homogeneously \citep[usually at low metallicity, see][]{Yoon2006a,Szecsi2015a}. In that case the star becomes extremely hot and compact, becoming a possible progenitor for long soft gamma-ray bursts \citep[and gravitational waves,][]{deMink2016a}.

\subsubsection{Evolution of the surface velocity}
\begin{figure}
\includegraphics[width=.6\columnwidth]{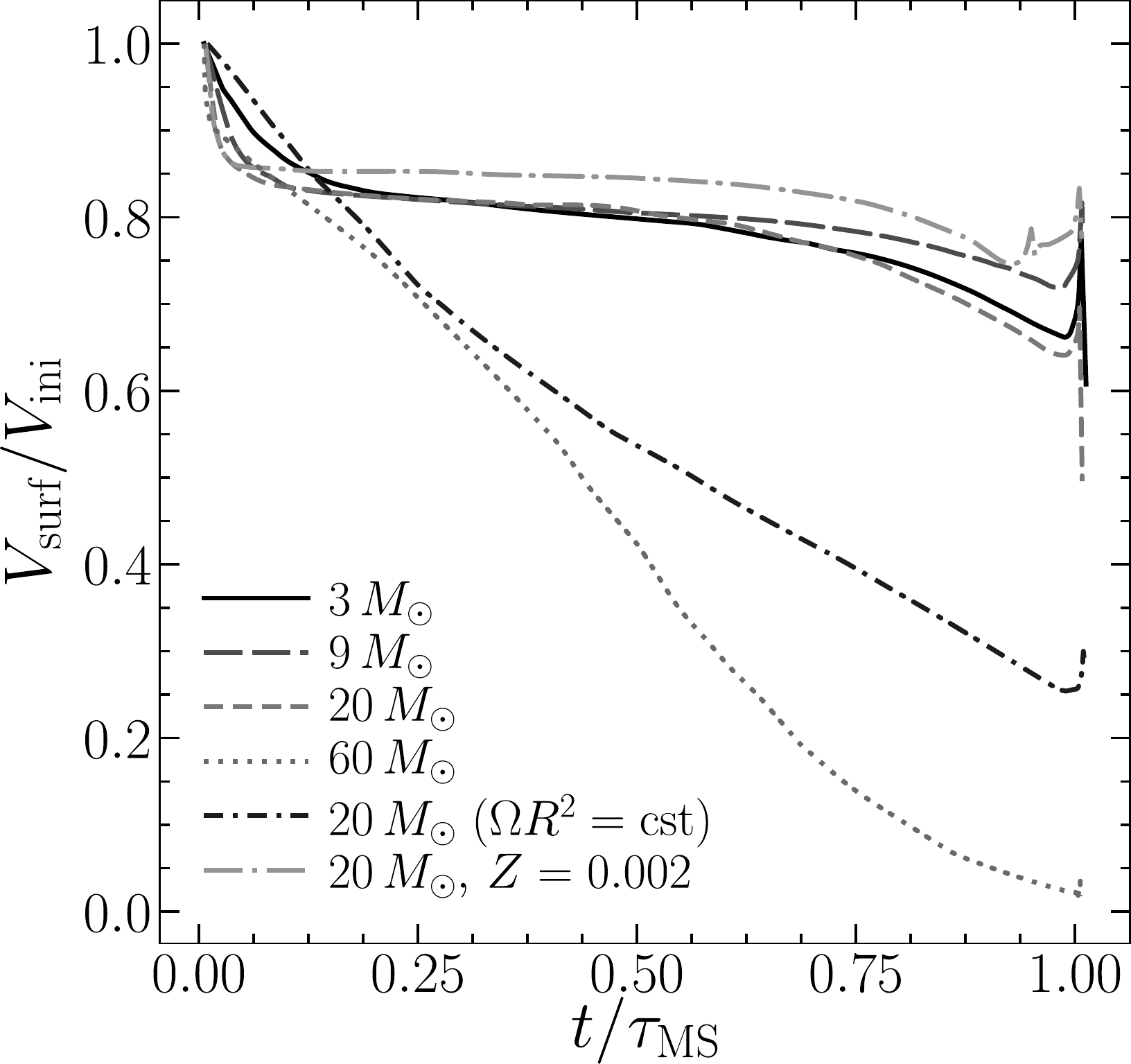}
\caption{Evolution of the equatorial velocity (normalised to the initial one) during the main sequence for various models at $Z=0.014$ computed with an initial rotation ratio $\omega=0.50$.}
\label{GeorgyFig:VeqEvol}
\end{figure}
If only the local conservation of angular momentum is considered, the surface velocity is expected to decrease during the main sequence, because of the natural expansion of the surface (see the short-dash-dotted curve in Fig.~\ref{GeorgyFig:VeqEvol} for a $20\,\msol$ model). On top of this natural evolution, two competing processes come into play: the mass loss (that removes angular momentum from the surface and triggers meridional currents), and the internal transport (that generally brings angular momentum from the centre to the surface). The strength of both the transport and the winds scales with the mass and metallicity, higher mass stars having stronger winds and more efficient transport, as do more metallic stars compared to low-$Z$ ones. The net result depends on their relative strength, as shown in Fig.~\ref{GeorgyFig:VeqEvol}. In the $3\,\msol$ and the $9\,\msol$ models (black solid and grey long-dashed curves respectively), the winds are negligible, the only difference is the strength of the transport, which is a little larger inside the $9\,\msol$ model. The $20\,\msol$ model (grey short-dashed curve) has a stronger transport, but the winds also become non-negligible, so the net result is a slight decrease of $V_\text{surf}$ during the main sequence. The case with no transport at all (local conservation of the angular momentum) is shown (dark grey, short-dashed-dotted curve) for comparison. In the $60\,\msol$ model, the transport is very strong but the wind effect largely dominates and the star is very efficiently braked. Actually in that mass range, any comparison of rotation rates with observations should be considered as a check for the mass-loss rate used rather than a check for the treatment of rotation itself. A low metallicity weakens the internal transport but also the winds, as illustrated by the grey long-dashed-dotted curve in Fig.~\ref{GeorgyFig:VeqEvol} ($20\,\msol$ model at $Z=0.002$), to be compared with the grey short-dashed one (same mass, $Z=0.014$).

Note that the surface velocity and its ratio to the critical velocity $V/V_\text{crit}$ evolve differently: $V_\text{crit}=\sqrt{\frac{2}{3}\frac{GM}{R_\text{pol}}}$ decreases always during the main sequence, because the mass decreases and the radius inflates. Therefore, even if the surface velocity decreases with the evolution, the ratio to the critical velocity usually increases during the main sequence.

\section{Conclusion}
\label{Georgy_conclu}

In this chapter, we have discussed some of the uncertainties inherent to the modelling of single massive stars. In the first section, we have shown how the inclusion of various mass-loss rates prescription can affect the evolution of massive stars. In particular, the mass loss during the red supergiant phase, which is still largely unknown, plays a key role for the subsequent evolution of the star and its final location in the Hertzsprung-Russell diagram.

Another key ingredient of stellar modelling is the treatment of convection. Here again, the difficulties of a correct treatment in classical 1d stellar evolution codes are a source of major shortcomings in our understanding of massive star life. The hopes for a better understanding and an improved modelling of convection come nowadays from multi-dimensional hydrodynamics simulations.

Finally, the implementation of rotation, which is suspected to play an important role as a mixing process inside massive stars, also suffers from uncertainties. The various possible ways of implementing rotation in stellar evolution codes make the prediction of simulations largely uncertain.

All of the uncertainties discussed above, related to single star evolution, also appear in the modelling of multiple systems, and they should be kept in mind when considering the results of any simulations involving stars.

\copyrightline{This material has been published in \textit{The Impact of Binaries on Stellar Evolution}, Beccari G. \& Boffin H.M.J. (Eds.).
This version is free to view and download for personal use only. Not for re-distribution, re-sale or use in derivative works. \copyright\ 2018 Cambridge University Press.}

\bibliography{GeorgyEkstrom_arXiv}
\bibliographystyle{cambridgeauthordate}

\end{document}